\documentclass[prl,longbibliography,twocolumn,superscriptaddress,amsmath,amssymb,verbatim]{revtex4-2}

\usepackage[english]{babel}
\usepackage[utf8]{inputenc}
\usepackage[skip=0.333\baselineskip]{caption}
\usepackage{subcaption}
\usepackage{graphicx}
\usepackage{subcaption}
\usepackage[colorinlistoftodos, color=green!40, prependcaption]{todonotes}
\usepackage{amsthm}
\usepackage{mathtools}
\usepackage{physics}
\usepackage{xcolor}
\usepackage{graphicx}
\usepackage[left=23mm,right=13mm,top=35mm,columnsep=15pt]{geometry} 
\usepackage{adjustbox}
\usepackage{placeins}
\usepackage[T1]{fontenc}
\usepackage{lipsum}
\usepackage{csquotes}
\usepackage{graphicx}
\usepackage{lmodern}
\usepackage{epstopdf}
\usepackage[colorlinks=true,citecolor=blue,linkcolor=blue,urlcolor=blue,bookmarks=true]{hyperref}
\usepackage[sort&compress]{natbib}
\usepackage[normalem]{ulem}

\begin{document}
\title{Anharmonic quantum effects of light particles in a spin-liquid material}

\author{F. Hotz}
\email{fabian.hotz@psi.ch}
\affiliation{PSI Center for Neutron and Muon Sciences CNM, 5232 Villigen PSI, Switzerland}

\author{M.\,Gomil\v{s}ek}
\email{matjaz.gomilsek@ijs.si}
\affiliation{Jo\v{z}ef Stefan Institute, Jamova cesta~39, SI-1000 Ljubljana, Slovenia}
\affiliation{Faculty of Mathematics and Physics, University of Ljubljana, Jadranska ulica~19, SI-1000 Ljubljana, Slovenia}

\author{T. Arh}
\affiliation{PSI Center for Neutron and Muon Sciences CNM, 5232 Villigen PSI, Switzerland}
\author{T. J. Hicken}
\affiliation{PSI Center for Neutron and Muon Sciences CNM, 5232 Villigen PSI, Switzerland}
\author{P. Umek}
\affiliation{Jo\v{z}ef Stefan Institute, Jamova cesta~39, SI-1000 Ljubljana, Slovenia}
\author{A. Zorko}
\affiliation{Jo\v{z}ef Stefan Institute, Jamova cesta~39, SI-1000 Ljubljana, Slovenia}
\affiliation{Faculty of Mathematics and Physics, University of Ljubljana, Jadranska ulica~19, SI-1000 Ljubljana, Slovenia}
\author{H. Luetkens}
\affiliation{PSI Center for Neutron and Muon Sciences CNM, 5232 Villigen PSI, Switzerland}






\date{\today} 

\begin{abstract}
The quantum behavior of light nuclei and other particles in materials challenges classical intuition and introduces novel phenomena. Here we demonstrate that muon spin spectroscopy ($\mu$SR) is a powerful tool for exploring the quantum effects of light particles, such as the muon, in condensed matter. The muon's quantum behavior is profoundly influenced by its surroundings, offering a unique proxy for also understanding the role of light nuclei and their influence on their local electronic environments. In Zn-barlowite, a candidate quantum spin liquid, we show that standard density functional theory (DFT) methods, which treat the muon as a classical point-like particle, fail to capture strong quantum anharmonic effects. Only by modeling the muon as a spatially extended quantum particle, thus accounting for the anharmonicity of its wavefunction, can the experimental $\mu$SR data be understood. This approach not only improves the interpretation of $\mu$SR results but also opens the door to studying the quantum effects of other light particles, like hydrogen and lithium nuclei, which can greatly influence material properties.
\end{abstract}


\maketitle
Materials containing light nuclei, such as hydrogen or lithium, play a key role in many important material classes, including superconductors, hydrogen-storage materials, semiconductors, Li-ion batteries and other advanced technologies \cite{markland2018nuclear,ranieri2022formation,ackland2017quantum,drozdov2015conventional,eremets2022high}. To understand these materials  requires a careful consideration of quantum mechanical effects, such as uncertainty in nuclear positions, a finite zero-point energy, and quantum tunneling, which can profoundly influence their properties. Accurately capturing these phenomena requires advanced computational methods that can address their complex interplay \cite{Markland2018,PhysRevLett.117.115702,Litman2024}. Therefore, there is a pressing need for approaches that seamlessly integrate nuclear quantum effects into $\textit{ab initio}$ descriptions of matter, paving the way for a deeper understanding of many, inherently quantum, materials. However, the development of accurate and efficient $\textit{ab initio}$ methods is hindered by the difficulty of experimentally determining the precise impact of these quantum effects.

Muon spin spectroscopy ($\mu$SR) offers a unique insight by being able to directly measure the quantum behavior of light particles in solid-state environments \cite{blundell2021muon,yaouanc2011muon,amato2024introduction}. In $\mu$SR, positive muons — charged (anti)particles ${\sim}9{\times}$ lighter than protons — are implanted into a material, where they interact with the surrounding magnetic and electronic fields. Their low mass enhances their sensitivity to quantum effects, by increasing their wavefunction spread (i.e., their quantum uncertainty in position), boosting zero-point energy (ZPE), and speeding up quantum tunneling. $\mu$SR enables direct measurement of these effects through dipolar interactions between the magnetic moment of the muon and those of the surrounding nuclei and electrons \cite{Moeller2013,blundell2023dft,PhysRevB.89.184425,PhysRevLett.114.017602}. Muons thus act as ideal proxies for studying the general quantum behavior of light particles including nuclei in materials. This can most clearly be seen in cases where classical, point-particle descriptions of muons fail to adequately describe experimental $\mu$SR data, obviating the need for a quantum description of muons \cite{gomilvsek2023many,PhysRevMaterials.3.073804,manas2021quantum}. By combining advanced \textit{ab initio} (density functional theory; DFT) and muon wavefunction calculations with $\mu$SR experiments, a much deeper understanding of light particles in matter can thus be achieved. We demonstrate this for the case of Zn-barlowite \cite{yuan2022emergence, PhysRevLett.128.157202, feng2017gapped,tustain2020magnetic}, where we find that the average muon positions are shifted from those of a classical muon by strong quantum anharmonic effects, while the stabilities of different candidate muon stopping sites are significantly impacted by quantum ZPE.
We confirm these \textit{ab initio} predictions by precision $\mu$SR experiments.

\begin{figure}[t!]
\centering
\includegraphics[width=9cm]{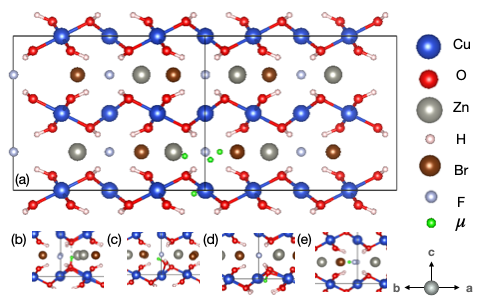}
\caption{(a) Candidate muon stopping sites in Zn-barlowite identified using classical DFT+$\mu$ (green) in the undistorted $2 \times 2 \times 1$ supercell. (b–d) Close-up views of the $\mu$–OH sites, showing local distortions induced by the muon. (e) F–$\mu$–Br stopping site with its associated local distortion. 
}
\label{classical_muon_position}
\end{figure}

We selected Zn-barlowite as the material for this study as it is a quantum spin liquid (QSL) candidate for which the exact knowledge of the muon's position and its interaction with its surroundings is of paramount importance for a correct interpretation of its magnetic properties extracted from $\mu$SR data \cite{khatua2023experimental}. Using DFT to simulate classical muons in a DFT+$\mu$ approach \cite{blundell2023dft} (see Supplementary Material for details) we predict four distinct classical muon stopping sites. The muon forms a $\mu$--OH complex at three muon sites [Figs. 1(b--d)] that are very close in energy (Table \ref{muon_energy_table}), while at the fourth site, the muon forms a F--$\mu$--Br complex [Fig.~\ref{classical_muon_position}(e)] with a classical energy 0.6 eV higher than at the lowest $\mu$--OH site. A large occupancy of this higher-energy site, was found in a prior experiment \cite{tustain2020magnetic}, however, this is in tension with the heuristic expectation that only sites with energies within approximately < 0.5 eV of the lowest energy site are typically occupied \cite{blundell2023dft,blundell2021muon,HUDDART2022108488}. 

The first and most accessible approximation to improve the classical point-particle description of DFT+$\mu$ is the quantum harmonic approximation, valid for small displacements of the muon from its classical position \cite{gomilvsek2023many}. Neglecting the quantum entanglement of muon and nuclear positions, the muon ZPE for each site can be estimated by summing the contributions $\hbar \omega_{j} / 2$ of the three main $\Gamma$-point normal phonon modes $\omega_{j}$ calculated by DFT with the muon embedded in the crystal \cite{RevModPhys.73.515,gomilvsek2023many}. 
This leads to total muon site energies listed in Table \ref{muon_energy_table}.\begin{table}[h]
\caption{\label{muon_energy_table}A comparison of classical and quantum (classical + ZPE) muon energies in the harmonic and anharmonic effective potentials for the four candidate muon stopping sites, relative to the lowest energy site ($\mu$--OH)$_1$. ${\Delta}r_\mu$ is the absolute quantum anharmonic shift of the average muon position from the classical position.}
\begin{ruledtabular}
\begin{tabular}{ccccc}
& ($\mu$--OH)$_{1}$&($\mu$--OH)$_{2}$ & ($\mu$--OH)$_{3}$ & F--$\mu$--Br\\
\hline
Classical (eV) & 0   & 0.0   &0.1  &0.6 \\
Harmonic (eV)   & 0.9 & 0.9 & 1 &1.3\\
Anharmonic (eV)  & 0.9 & 1 & 1.1 &1.2\\
$\Delta r_\mu~(\textit{\r{A}})$   & 0.17& 0.12 & 0.08 &0.06\\
\end{tabular}
\end{ruledtabular}
\end{table} The relative stability of the highest-energy F--$\mu$--Br site is improved from the classical calculations by taking into account the quantum ZPE. However, since at this site the muon is shared between strongly electronegative F$^{-1}$ and Br$^{-1}$ ions, the potential could be strongly anharmonic, making the harmonic approximation insufficient. Furthermore, as the highest-frequency normal mode at the $\mu$--OH sites is along the $\mu$--O direction, one could expect strong anharmonic effects also at $\mu$--OH sites, with a higher energy penalty for displacements towards the oxygen than away from it due to the positive muons being electrostatically repulsed from the positive nucleus of oxygen.

\begin{figure}[t]
\centering
\includegraphics[width=8cm]{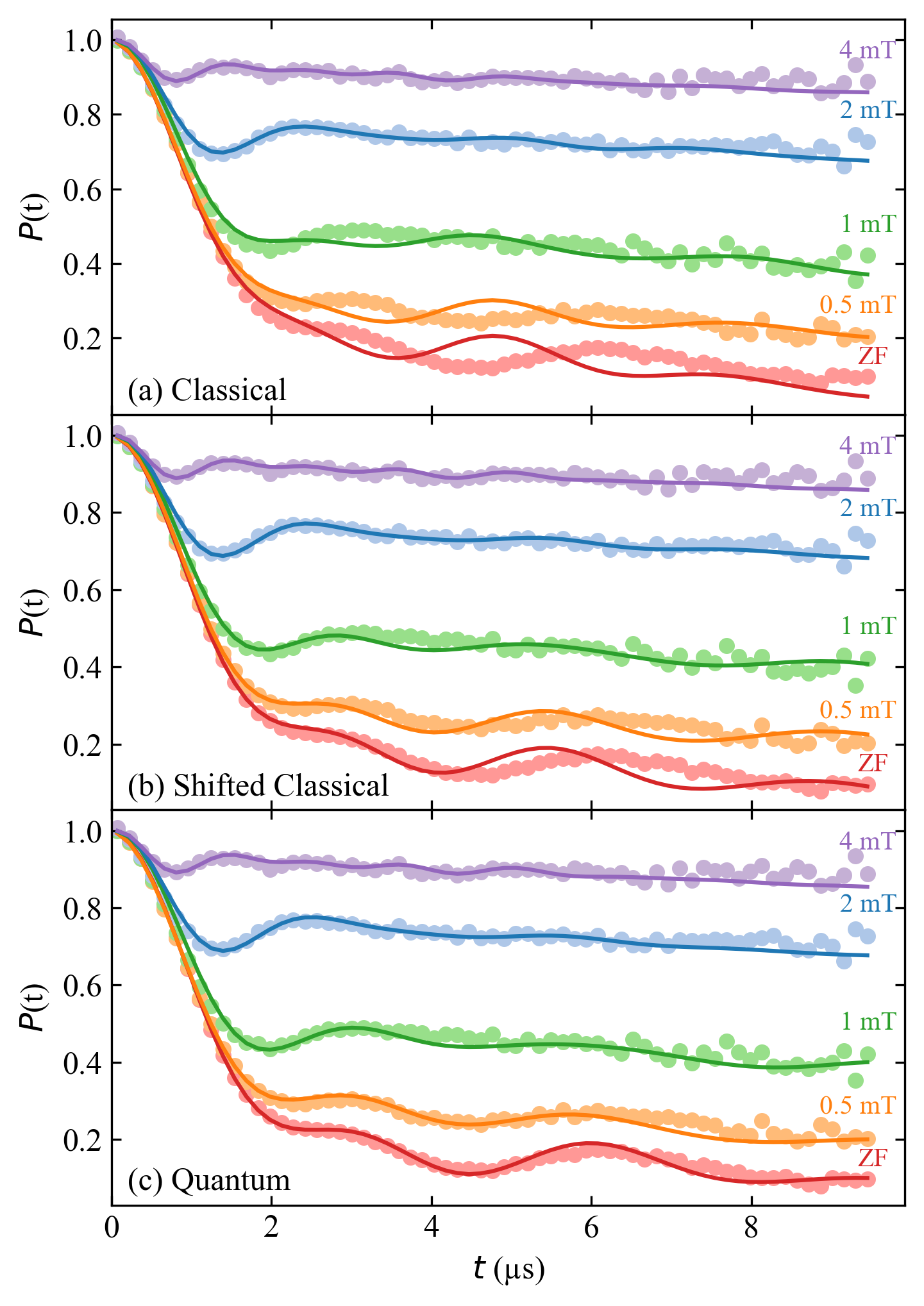}
\caption{
Experimental LF decoupling $\mu$SR data on Zn-barlowite (points) at 20 K, fitted using Eq. (1) (curves) assuming (a) a classical muon position, (b) a point-particle muon shifted to the average quantum muon position (Fig. 3), and (c) a fully quantum muon. 
}
\label{ZF_LF_DFT}
\end{figure}

 To distinguish between the harmonic and anharmonic scenarios, we turn to experiment. The aim is to determine the unique nuclear neighborhoods and muon wavefunctions at sites F--$\mu$--Br and $\mu$--OH in Zn-barlowite, exploiting the coupling between muon and nuclear spins \cite{blundell2021muon,LORD2006472,lord2000muon}. For this purpose, longitudinal field (LF) $\mu$SR experiments were performed on Zn-barlowite powder samples on the FLexible Advanced MuSR Environment (FLAME) instrument at PSI, Switzerland, in fields up to $4$~mT (Fig. \ref{ZF_LF_DFT}). 

To check whether the predicted muon stopping sites account well for experimental data we compute the time evolution of the muon polarization due to dipolar interactions with nuclear spins at each of the four muon sites $i$ via exact-diagonalization calculations, giving
$P_{i}^{\text{stat}}(t) = \left<\frac{1}{2} \text{Tr}\left \{ \sigma^{\mu}_{\hat{n}} \exp(\frac{i}{\hbar} \hat{\mathcal{H}}_{i} t)\sigma^{\mu}_{\hat{n}} \exp(\frac{-i}{\hbar} \hat{\mathcal{H}}_{i}t)  \right \} \right>_{\hat{n}}$,  where $\sigma^{\mu}_{\hat{n}}$, is the muon's Pauli spin operator in the detector direction $\hat{n}$ relative to a powder grain, and $\langle \cdots \rangle_{\hat{n}}$ represents a powder average \cite{blundell2021muon, Wilkinson2020, PhysRevLett.56.2720,lord2000muon}. The Hamiltonian $\hat{\mathcal{H}}_{i}$ consists of the Zeeman interaction due to the external applied field $B$, dipolar interactions between the muon and nuclear spins, and quadrupole interaction of the Br nuclei (spin $I$ = 3/2) with the local electric field gradient \cite{gomilvsek2023many, PhysRevLett.129.097205}.
The latter two terms depend on the muon position, providing experimental sensitivity to the muon position. Furthermore, the electron spins of magnetic Cu$^{2+}$ ions remain disordered and fluctuate in the QSL ground state of Zn-barlowite \cite{yuan2022emergence,Wang_2021,feng2017gapped,tustain2020magnetic}, further dynamically relaxing the muon's spin. Due to these fluctuations, the final muon polarization at a given muon site can be written as \cite{amato2024introduction,blundell2021muon} $P_i (t) =  P^{\text{stat}}_i(t)P^{\text{dyn}}(t)$. Here, the contribution due to electron spin fluctuations is accounted for by the exponential function $P^{\text{dyn}}(t) = \text{exp}(-\lambda_{\text{dyn}} t)$, where $\lambda_{\text{dyn}}$ is the dynamical relaxation rate, which depends on the applied field $B$. Although each local surrounding could have a different relaxation rate $\lambda_{\text{dyn}}$ the final fits are not noticeably improved by allowing this, so we constrain $\lambda_{\text{dyn}}$ to be equal at all sites to minimize the number of free parameters. 

Adding the contributions from all muon sites together, we obtain the final fitting function for the muon polarization 
\begin{align}
\label{experimental_polarization_fkt_ext_B}
P(t) = f_s \sum_{i=0}^4 f_i P_i(t) + (1 - f_s) P_\mathrm{bgd}(t) ,
\end{align}
where $f_s$ is the fraction of muons that stop in the sample and $f_i$ is the fraction of those muons that end up at a site $i$ with polarization $P_i(t)$ (see above), where $i = 0$ denotes the F--$\mu$--Br site and $i = 1, 2, 3$ to the three ($\mu$--OH)$_i$ sites, respectively. $P_\mathrm{bgd}(t)$ is the LF Gaussian Kubo--Toyabe function under an applied field \cite{PhysRevB.20.850}, which describes the background contribution of muons that stop outside of the sample.


 Fit results using Eq. (\ref{experimental_polarization_fkt_ext_B}) for a muon treated as a classical point particle at the classical DFT+$\mu$ position are shown in Fig.~\ref{ZF_LF_DFT}(a). These yield a reduced $\chi^{2}$/DOF = 9.8, showing that the classical prediction fails to reproduce the experiment. The discrepancy is especially prominent at low applied fields. To improve the fit accuracy we need to model the muon wavefunction. To incorporate anharmonic muon effects,
\begin{figure*}[t]
  \centering
  \begin{subfigure}{.475\linewidth}
    \includegraphics[width=\linewidth]{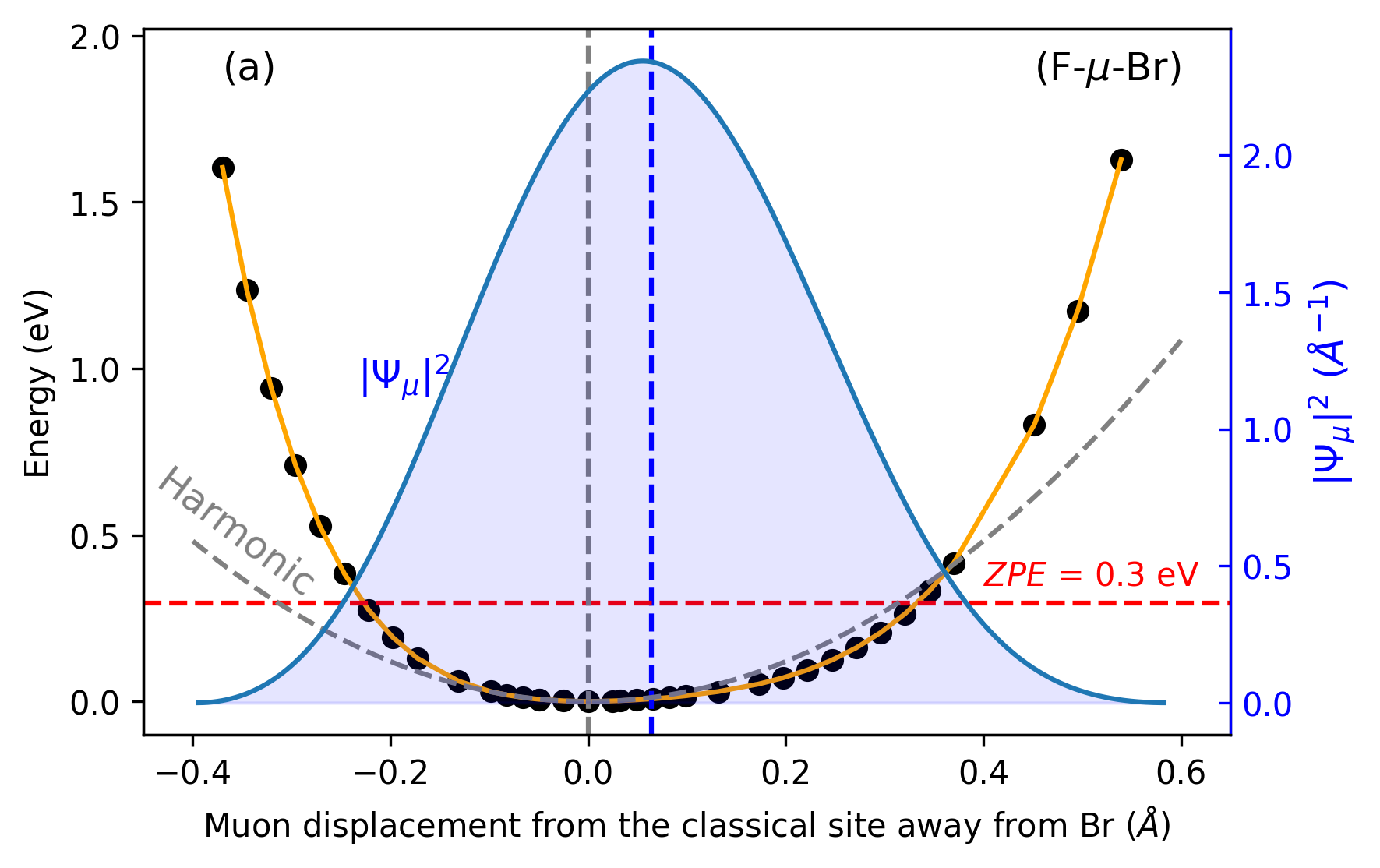}
    \label{MLEDdet}
  \end{subfigure}\hfill
  \begin{subfigure}{.475\linewidth}
    \includegraphics[width=\linewidth]{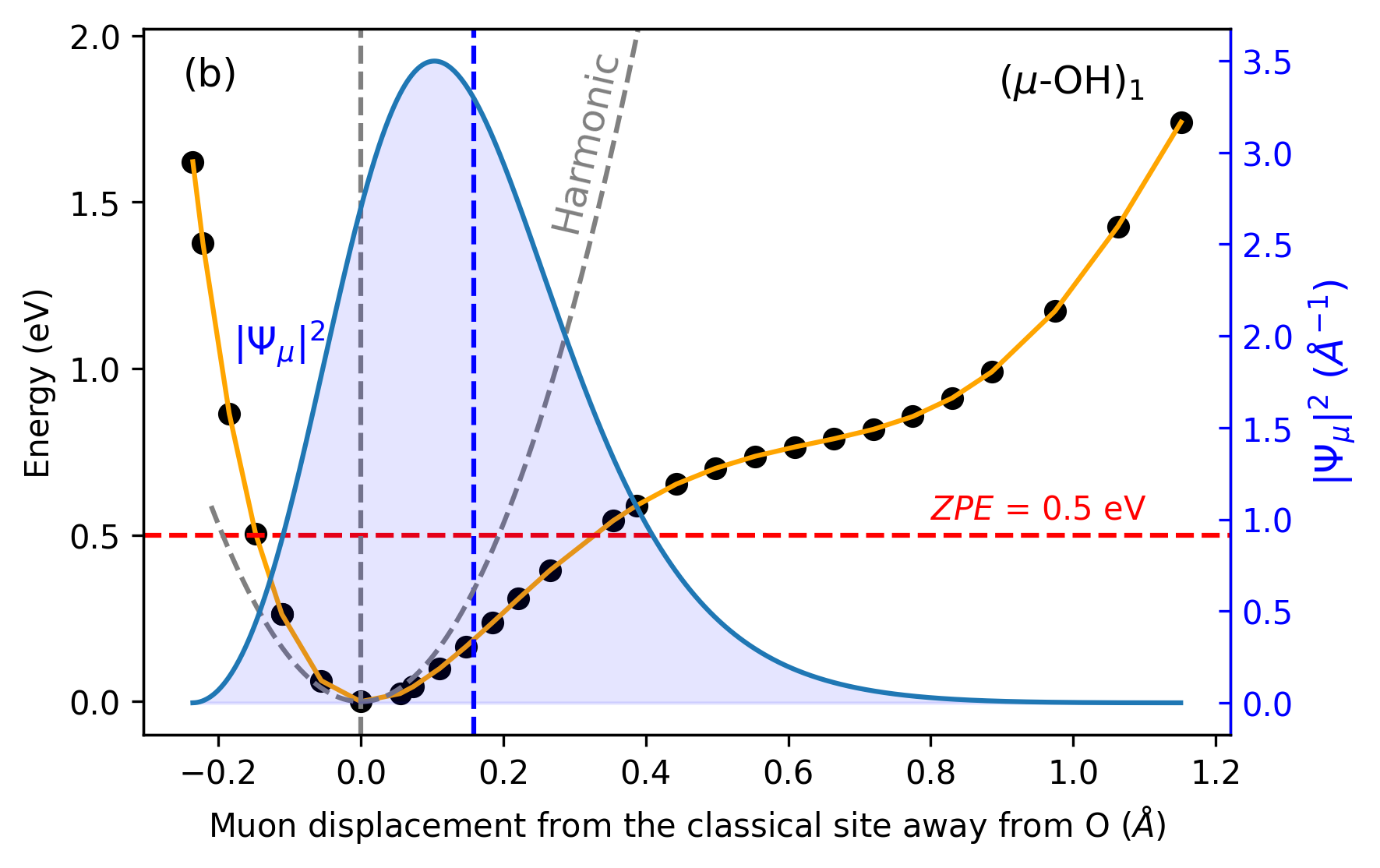}
    \label{energydetPSK}
  \end{subfigure}
  \medskip
  \begin{subfigure}{.475\linewidth}
    \includegraphics[width=\linewidth]{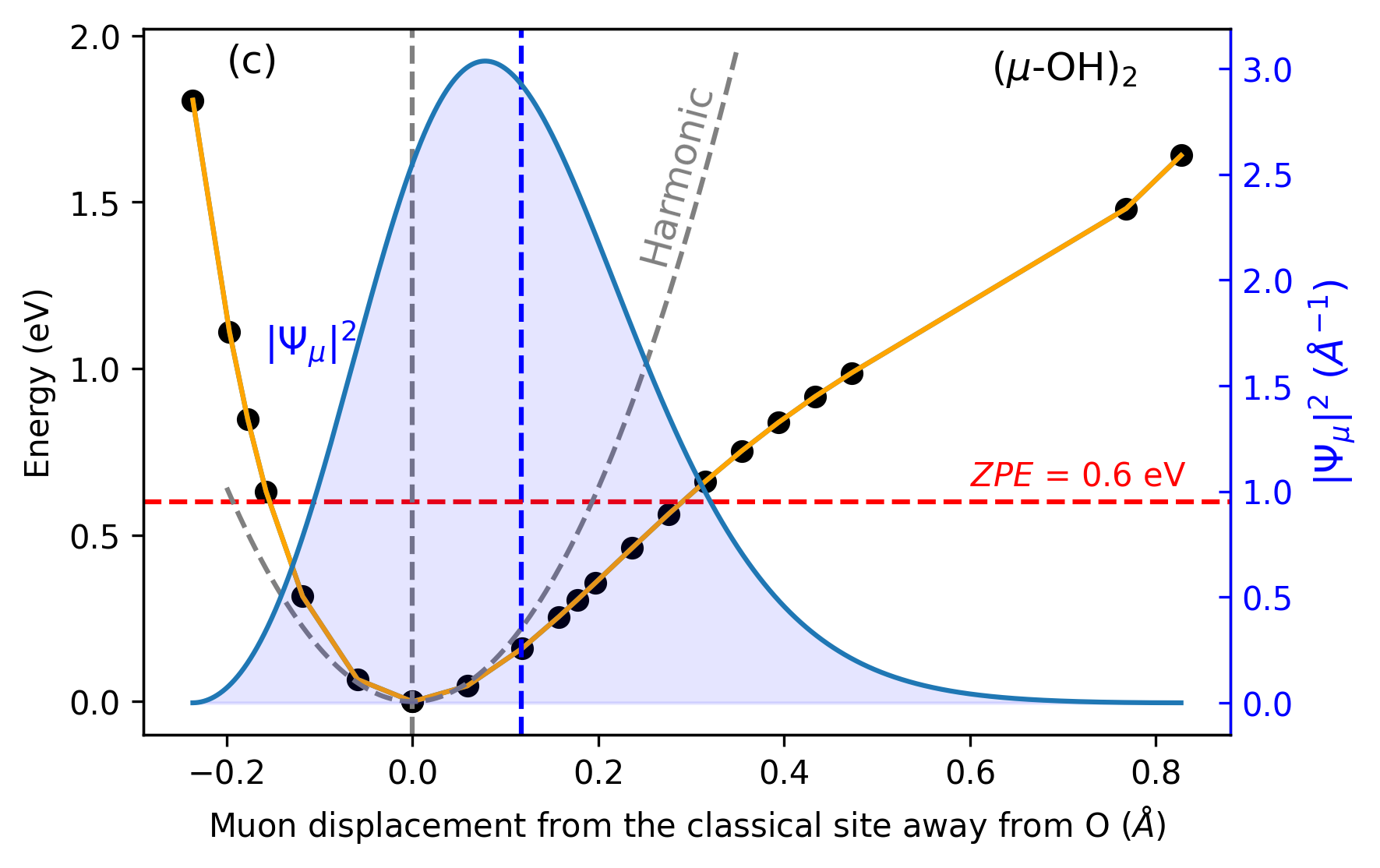}
    \label{velcomp}
  \end{subfigure}\hfill
  \begin{subfigure}{.475\linewidth}
    \includegraphics[width=\linewidth]{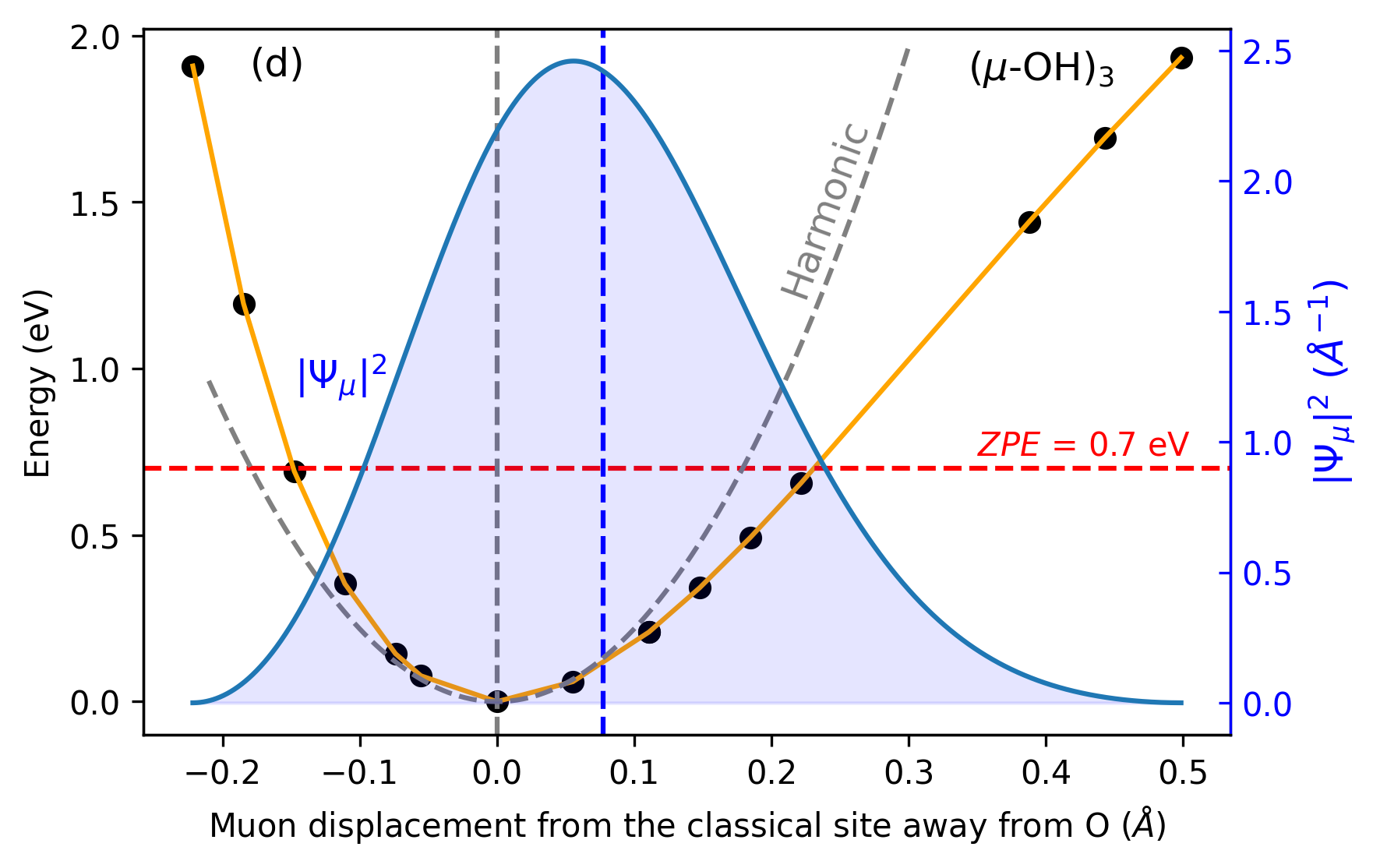}
    \label{estcomp}
  \end{subfigure}
  \caption{Effective potentials in the weakly-bound adiabatic limit (relative to the classical energy) for muon displacements along the (a) F--Br and (b--d) ($\mu$--O)$_i$ directions at F--$\mu$--Br and $\mu$--OH sites, respectively, from DFT (black points, interpolated by orange curves). These are compared to the effective potential under the harmonic approximation (dashed grey). Also shown here, the quantum muon probability distributions $|\psi_{\mu}|^{2}$ in the adiabatic potentials (blue curves) with the resulting average muon displacements ${\Delta}r_\mu$ (dashed vertical blue lines) away from their classical/harmonic positions (dashed vertical gray lines).}
  \label{anharmonic}
\end{figure*}
we have calculated the effective (adiabatic) anharmonic potential for the muon around each candidate stopping site under the assumption of negligible muon--nuclear entanglement in position (giving a single-particle, weakly-bound adiabatic potential \cite{gomilvsek2023many}), and assuming a separable potential for the muon with eigenaxes oriented along the normal mode directions obtained from the harmonic phonon DFT calculation. The effective potentials along the most anharmonic directions at the four muon stopping sites are shown in Fig.~\ref{anharmonic}, while the potential along the perpendicular directions at these sites are found to be nearly-harmonic (see the Supplementary Material). 
Along each normal direction, the resulting 1D Schr\"odinger equation for the muon wavefunction was solved numerically using a finite-difference method. At the F--$\mu$--Br site, the potential along the F--Br direction is flatter than expected from the harmonic approximation [Fig.~\ref{anharmonic}(a)],  reducing the muon ZPE and stabilizing this site [see Table $\ref{muon_energy_table}$]. Since the muon is located between two negatively charged ions, the average muon position is shifted by ${\Delta}r_\mu$ $\approx0.06$ $\mathring{\text{A}}$ away from the Br$^{-}$ compared to the classical site. On the other hand, at $\mu$--OH sites the adiabatic potentials along the $\mu$--O directions are highly anharmonic, shifting the average muon position by up to 0.17 $\mathring{\text{A}}$ [Fig (\ref{anharmonic})(b--d) and Table \ref{muon_energy_table}], which represents a significant fraction of the expected 1 $\mathring{\text{A}}$ $\mu$--O distance for a classical muon.

\noindent The fits to the $\mu$SR spectra can already be improved by shifting a point-particle muon, from the classical to the average quantum position, which is anharmonically shifted away from the classical one by ${\Delta}r_\mu$ (Table \ref{muon_energy_table}), as shown in Fig.~\ref{ZF_LF_DFT}(b). This yields $\chi^2/\mathrm{DOF} = 3.5$. However, the fit quality is further substantially improved by taking into account the quantum uncertainty in muon position; in other words, using its full computed wavefunction $\psi_\mu$ to construct an effective muon--nuclear spin hamiltonian, by averaging point-particle dipolar muon--nuclear hamiltonians over the muon probability distribution $|\psi_{\mu}|^{2}$ \cite{gomilvsek2023many}. The result is shown in Fig.~\ref{ZF_LF_DFT}(c), which yields the final reduced $\chi^2$/DOF = 1.9. 
This good agreement between experimental data and calculation validates our fully quantum approach to an $\textit{ab initio}$ treatment of implanted muons. Our results show that to determine the muon stopping sites in compounds with strongly electronegative ions like F$^-$, Cl$^-$, and Br$^-$, or functional groups, like OH$^-$, it  is necessary to treat the muon as a quantum particle in an anharmonic potential.
Additionally, this approach provides a way towards understanding the quantum anharmonic effects of other light particles, such as hydrogen or lithium nuclei, in materials. Using $\mu$SR measurements of muons as proxy particles allows us to probe these effects in an extreme quantum limit, as the muon’s low mass amplifies these quantum phenomena. This makes $\mu$SR a powerful tool to investigate the quantum behavior of light particles in lattice environments  and allows us to quantitatively verify the quantum mechanical calculations of these effects.

In conclusion, our study demonstrates that quantum effects for light particles, such as muons, can profoundly influence experimental observables in materials. Specifically, we have demonstrated a pronounced anharmonic shift in the average muon position, the impact of quantum uncertainty of the muon position on the observed $\mu$SR spectra, and the stabilization of a higher-energy muon site due to it having a lower quantum zero-point energy than other candidate muon sites. $\mu$SR provides a unique experimental opportunity to use the muon as an effective model system to study the quantum behavior of light nuclei. Our study underscores the critical need to incorporate the quantum effects of light particles and nuclei into $\textit{ab initio}$ descriptions of materials, which are essential for accurate predictions and a deeper understanding of material properties $\cite{markland2018nuclear}$.

FH acknowledges the financial support of the Swiss National Science Foundation through Program No. 192109. Experimental work was carried out at the Swiss Muon Source, PSI, Switzerland.
The MERLIN HPC cluster, PSI, provided computing resources.
Computing resources were also provided by the STFC Scientific Computing Department’s SCARF cluster. MG and AZ acknowledge the financial support of the Slovenian Research and Innovation Agency through Program No. P1-0125 and Projects No. Z1-1852, N1-0148, J1-2461, J1-50008, J1-50012, and N1-0356, N1-0345, and J2-60034.


\newpage
\appendix
\onecolumngrid
\begin{center}
{\large \bf Supplementary Information:\\
Anharmonic quantum effects of light particles in a spin liquid material}\\
\vspace{0.3cm}
F. Hotz,$^{1,\,\ast}$, M. Gomil\v sek,$^{2,\,3,\,\ast}$, T. Arh,$^{1}$, T.J. Hicken,${^1}$, P. Umek,$^{1}$ and A. Zorko$^{2,\,3}$, H. Luetkens,$^{1}$ \\
$^1${\it PSI Center for Neutron and Muon Sciences CNM,  CH-5232 Villigen PSI, Switzerland}\\
$^2${\it Jo\v{z}ef Stefan Institute, Jamova c.~39, SI-1000 Ljubljana, Slovenia} \\
$^3${\it Faculty of Mathematics and Physics, University of Ljubljana, Jadranska u.~19, SI-1000 Ljubljana, Slovenia}\\
\end{center}

\twocolumngrid

\section{Classical muon stopping sites}
\label{geom_appendix}
Via a DFT+$\mu$ approach we have identified the classical muon stopping sites in Zn-barlowite \cite{Moeller2013,blundell2023dft} using the plane-wave DFT code CASTEP \cite{ClarkSegallPickardHasnipProbertRefsonPayne+2005+567+570}. Using the MuFinder software  \cite{HUDDART2022108488}, 32 initial muon candidates were put in low-symmetry positions, and after converging the geometry optimization below a 0.01 eV/$\AA$ force tolerance, the sites were clustered into four final candidate sites using connected-component clustering \cite{HUDDART2022108488}. The calculations were performed in a supercell containing $2 \times 2 \times 1$ conventional unit cells of Zn-Barlowite \cite{BJORKMAN20111183} using the standard PBEsol functional \cite{PhysRevLett.100.136406}, an energy cutoff of 2000 eV, and a $2 \times 2 \times 3$ Monkhorst-Pack grid \cite{PhysRevB.13.5188} for $k$-point sampling. The relatively high convergence parameters were used to ensure well-converged phonon calculations with the force converged to 10$^{-4}$~eV/$\AA$, well below the force tolerance of geometry optimization. The +1 charge state of the positive muon, which was modeled by an ultrasoft hydrogen pseudopotential with a reduced mass and an enhanced gyromagnetic ratio, was modeled by a +1 charge of the supercell. All calculations were performed in a spin-polarized mode with an effective Hubbard $U_\mathrm{eff} = U - J = 5$~eV, where $U$ is the bare Hubbard interaction strength and $J$ is Hund's coupling, consistently with previous studies \cite{Hubbard_U, PhysRevB.94.125136,PhysRevB.92.094417}. To verify the reliability of our simulations, we have recalculated the coordinates of the converged F--$\mu$--Br site and the lowest energy ($\mu$--OH)$_{1}$ site with different functionals and compared their values in Table \ref{functional_comparision}. We have compared the local density approximation (LDA) functional \cite{PhysRev.140.A1133} and the PBE functional \cite{perdew1996generalized}  to the PBEsol functional used in the rest of the paper.

\begin{table}[h]
\caption{\label{functional_comparision}The $x$, $y$, and $z$ fractional coordinates of the muon stopping sites in the $2 \times 2 \times 1$ conventional unit cell for three different DFT functionals: LDA, PBE, and PBEsol. For all three functionals, a Hubbard $U_\mathrm{eff}$ = 5 eV is used. The change in functionals leads to almost no change in the coordinates for the lowest-energy ($\mu$--OH)$_{1}$ site, but a significant change in the $y$-coordinate of the  F--$\mu$--Br site.}
\begin{ruledtabular}
\begin{tabular}{cccc}
 & LDA &PBE & PBEsol\\
 \hline
F--$\mu$--Br(x)  & 0.0747   & 0.0884 & 0.0733 \\
 F--$\mu$--Br(y)  & 0.1533 & 0.1806 & 0.1506\\
  F--$\mu$--Br(z) & 0.2488 & 0.2489 & 0.2489\\
 ($\mu$--OH)$_{1}$(x)    & 0.214& 0.212 &  0.213\\
 ($\mu$--OH)$_{1}$(y) &  0.109& 0.1088  & 0.109  \\
 ($\mu$--OH)$_{1}$(z) & 0.221& 0.222& 0.222\\
\end{tabular}
\end{ruledtabular}
\end{table}

\section{Zero-point energy in the harmonic and adiabatic approximations}
\label{quantum_muon_appendix}
\begin{figure*}[h]
  \centering
  \begin{subfigure}{.43\linewidth}
    \includegraphics[width=\linewidth]{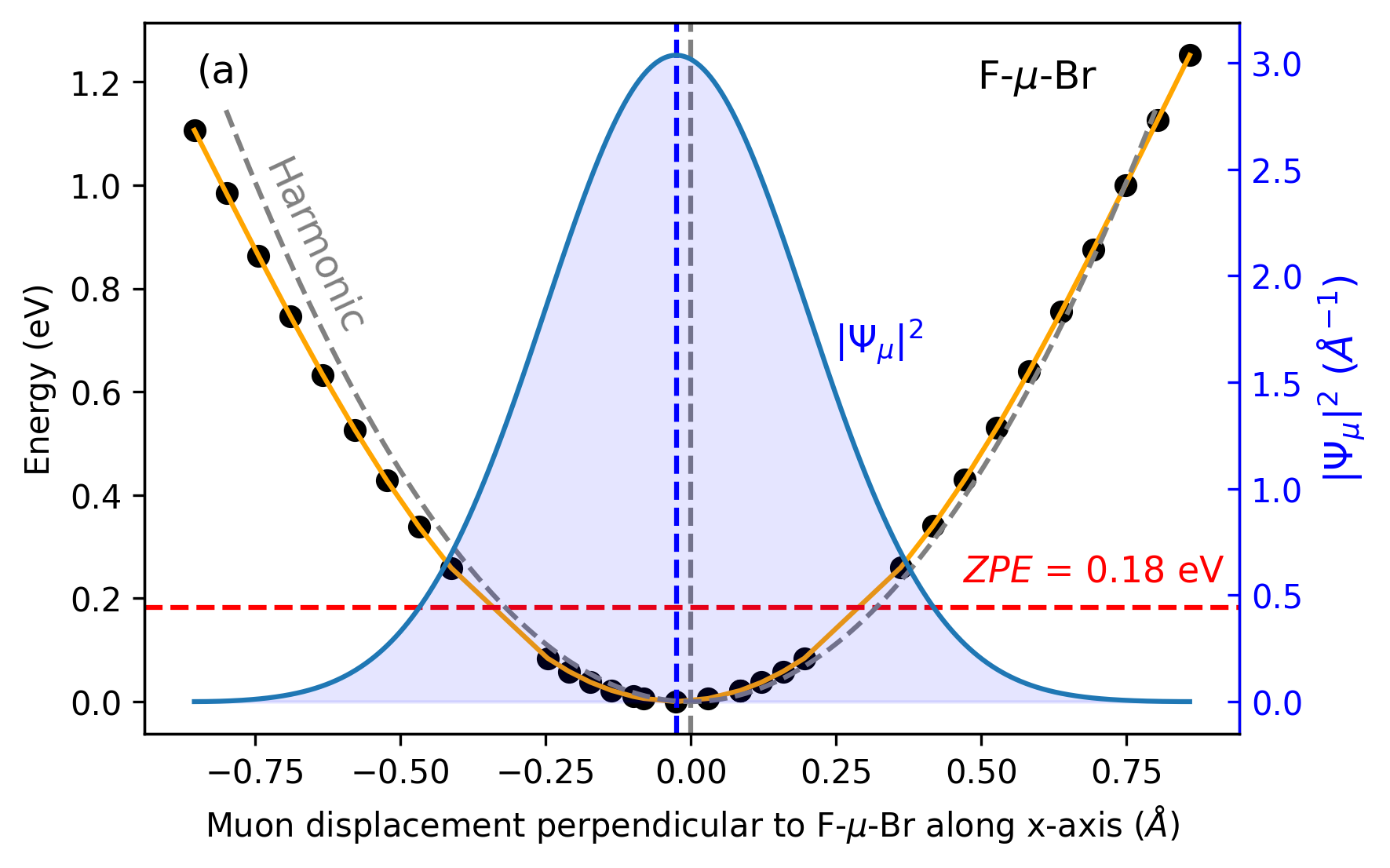}
    \label{ZPE_FBr_X}
  \end{subfigure}\hfill
  \begin{subfigure}{.43\linewidth}
    \includegraphics[width=\linewidth]{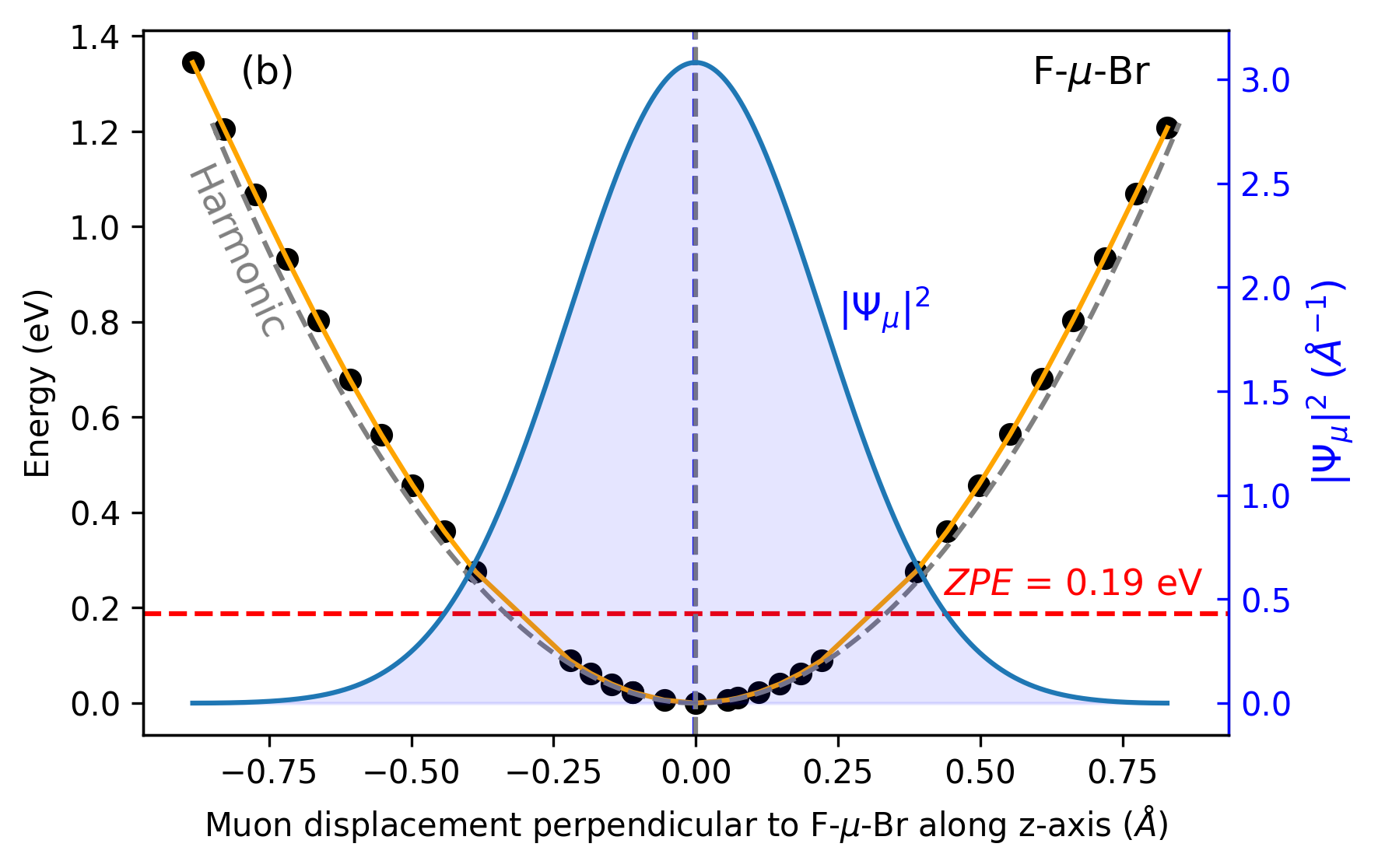}
    \label{ZPE_FBr_Z}
  \end{subfigure}
  \medskip
  \begin{subfigure}{.43\linewidth}
    \includegraphics[width=\linewidth]{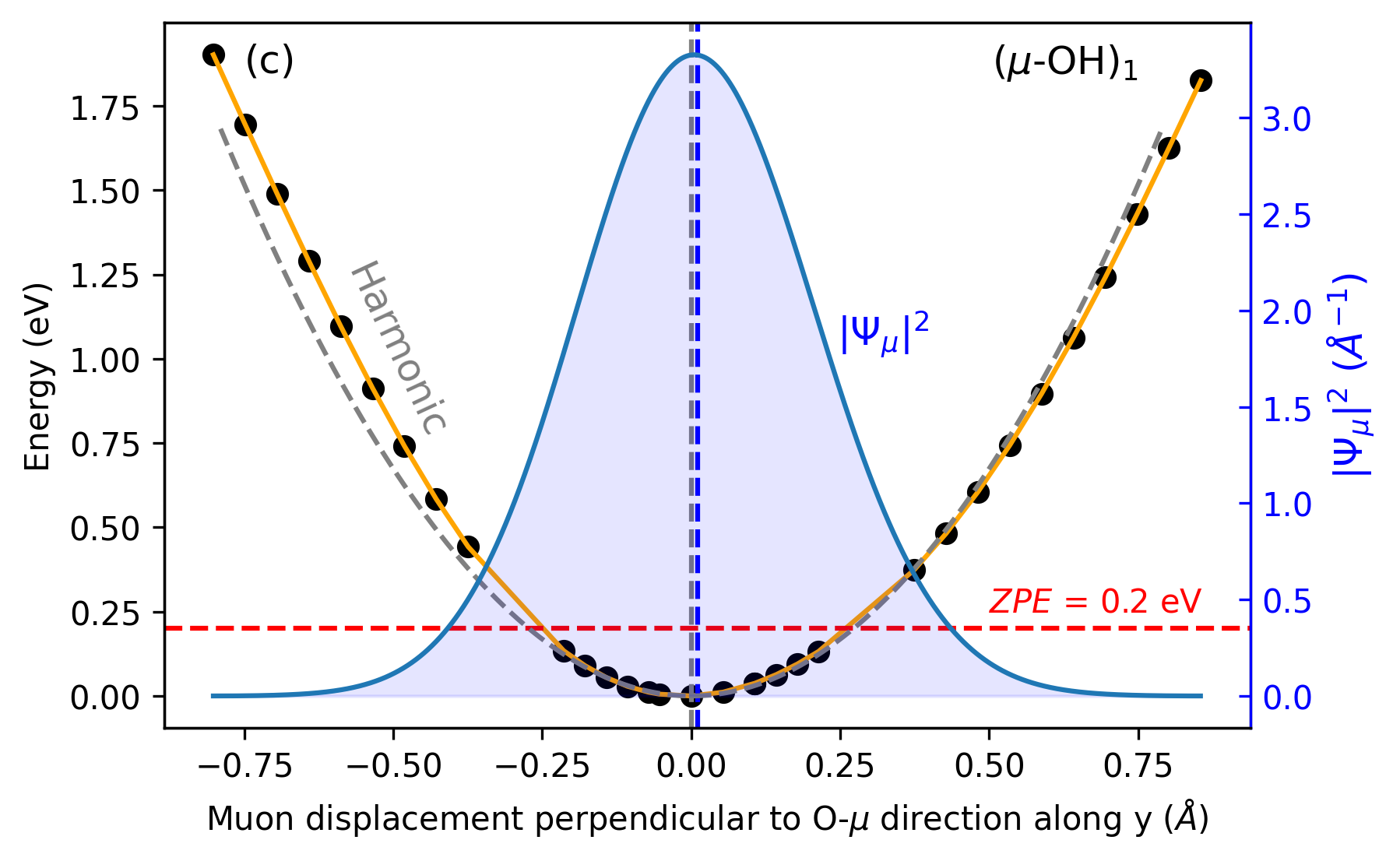}
    \label{ZPE_site5_Y}
  \end{subfigure}\hfill
  \begin{subfigure}{.43\linewidth}
    \includegraphics[width=\linewidth]{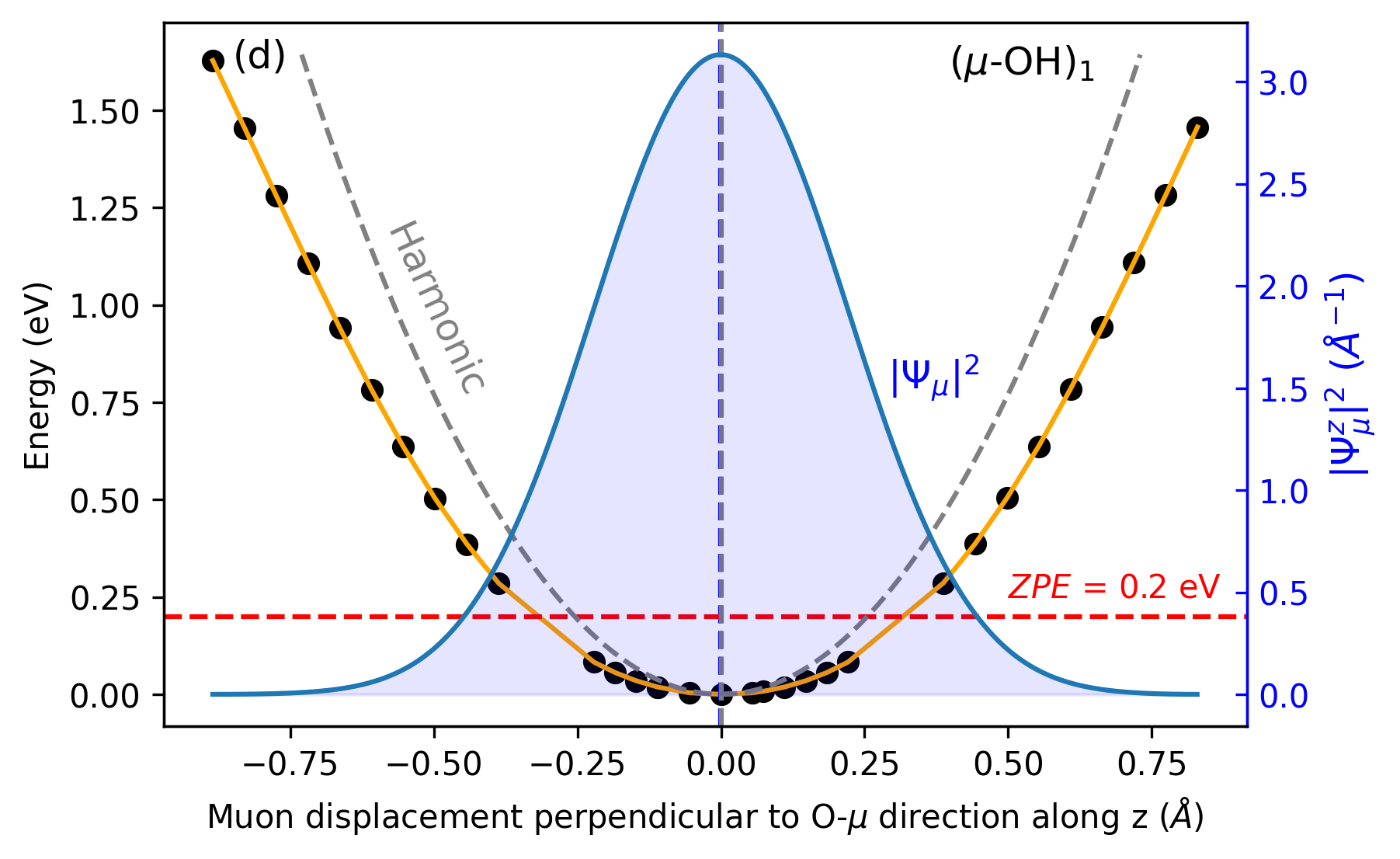}
    \label{ZPE_site5_Z}
  \end{subfigure}
  \medskip
  \begin{subfigure}{.43\linewidth}
    \includegraphics[width=\linewidth]{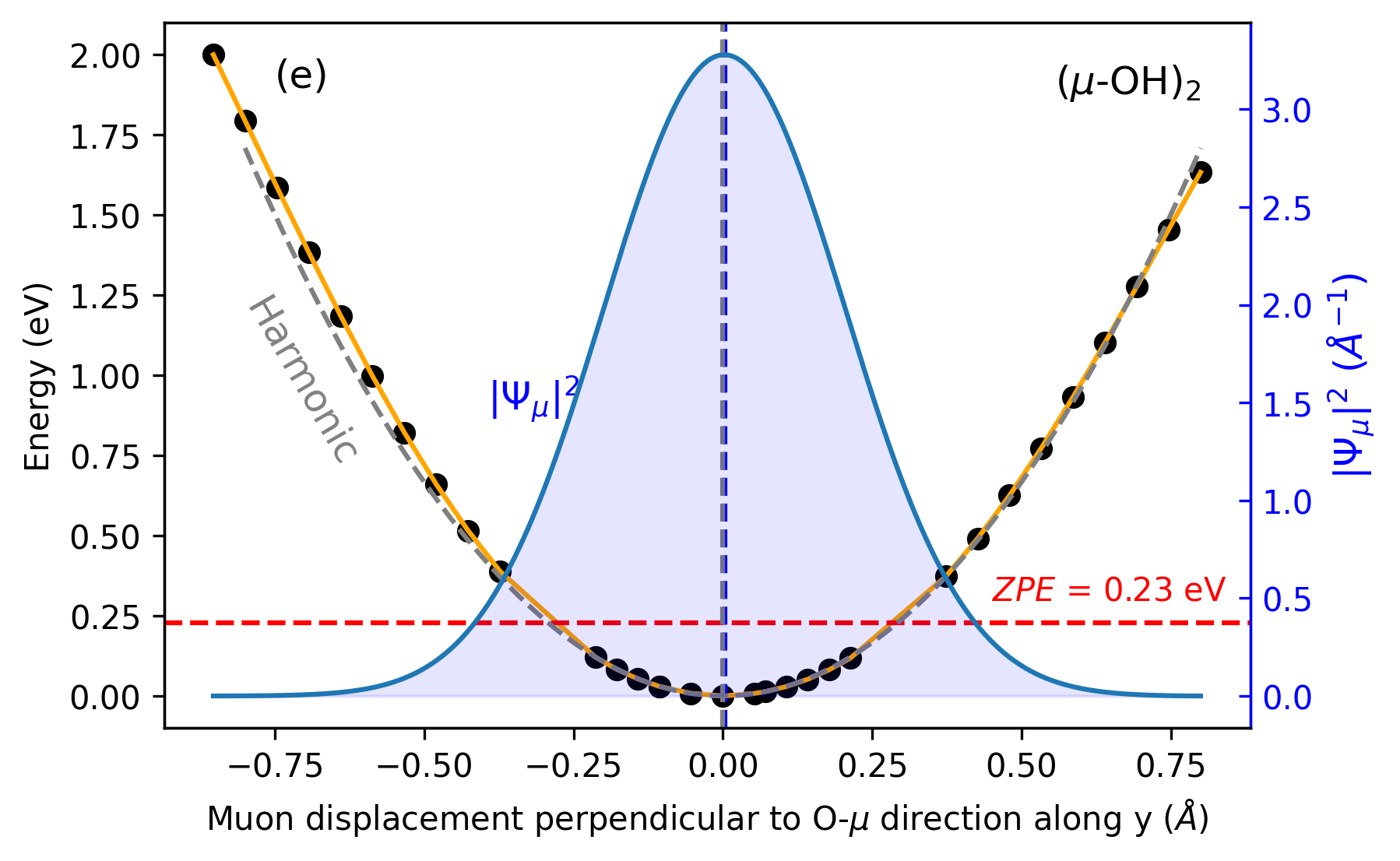}
    \label{ZPE_site6_Y}
  \end{subfigure}\hfill
  \begin{subfigure}{.43\linewidth}
    \includegraphics[width=\linewidth]{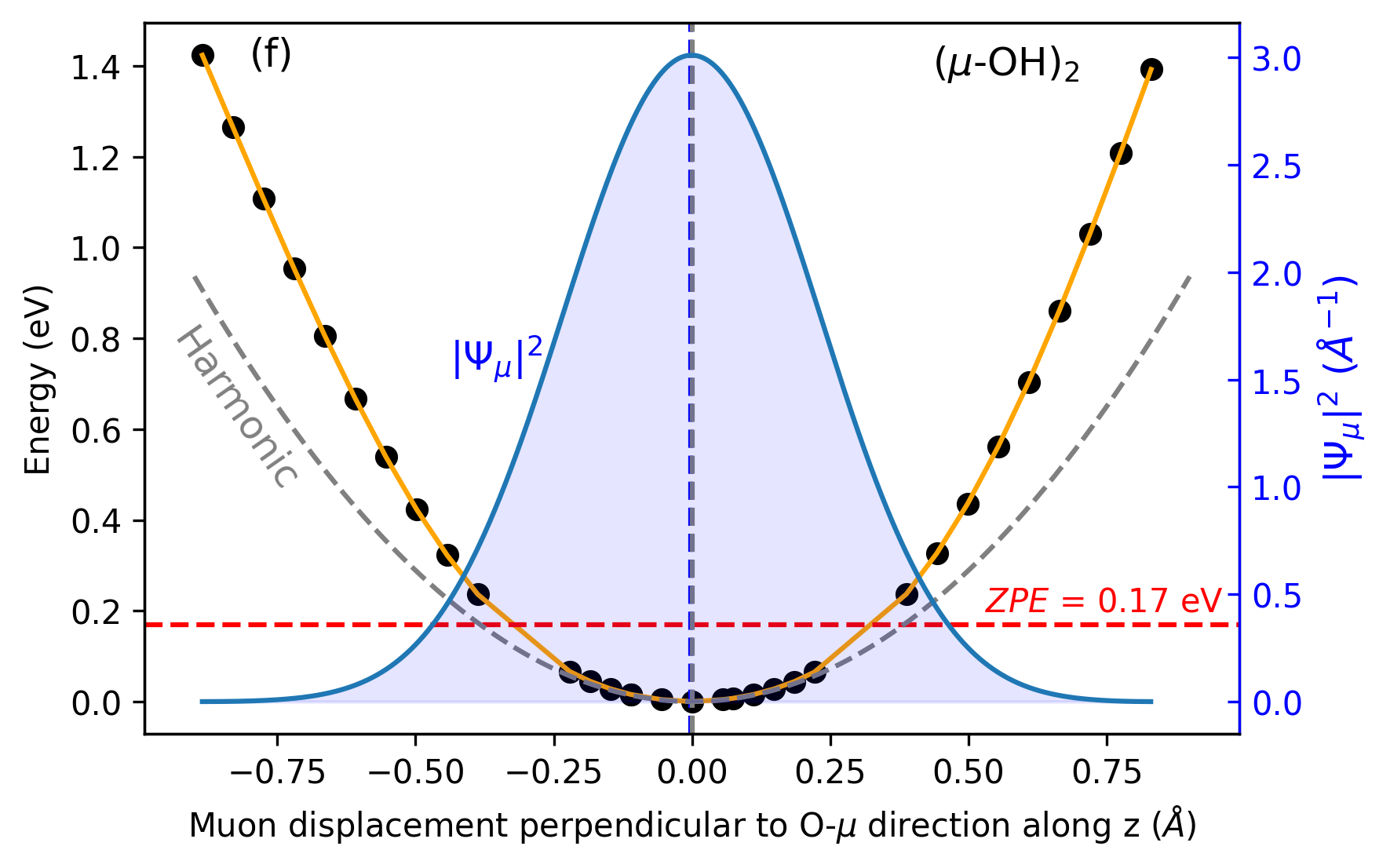}
    \label{ZPE_site6_Z}
  \end{subfigure}
    \medskip
  \begin{subfigure}{.43\linewidth}
    \includegraphics[width=\linewidth]{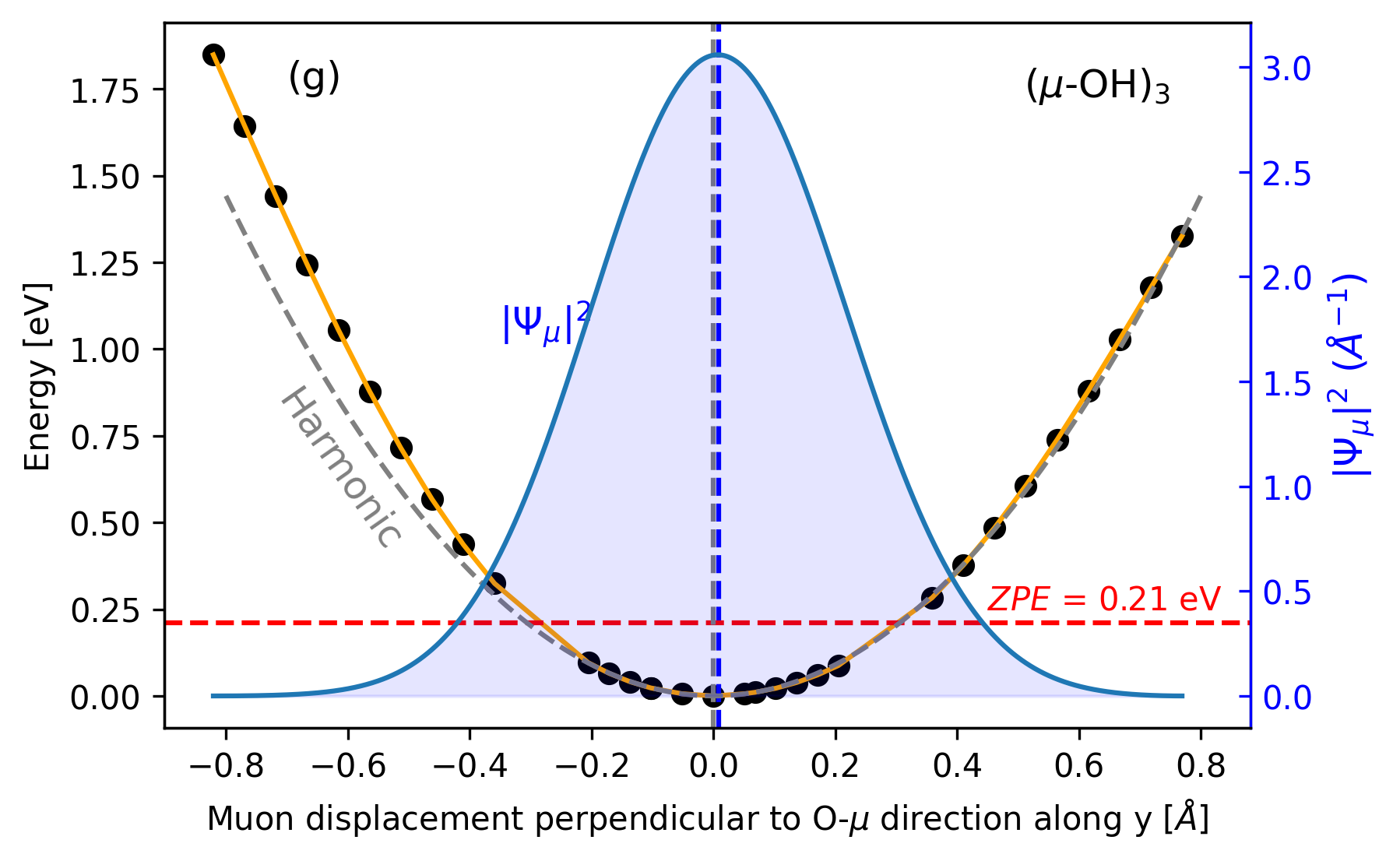}
    \label{ZPE_site4_Y}
  \end{subfigure}\hfill
  \begin{subfigure}{.43\linewidth}
    \includegraphics[width=\linewidth]{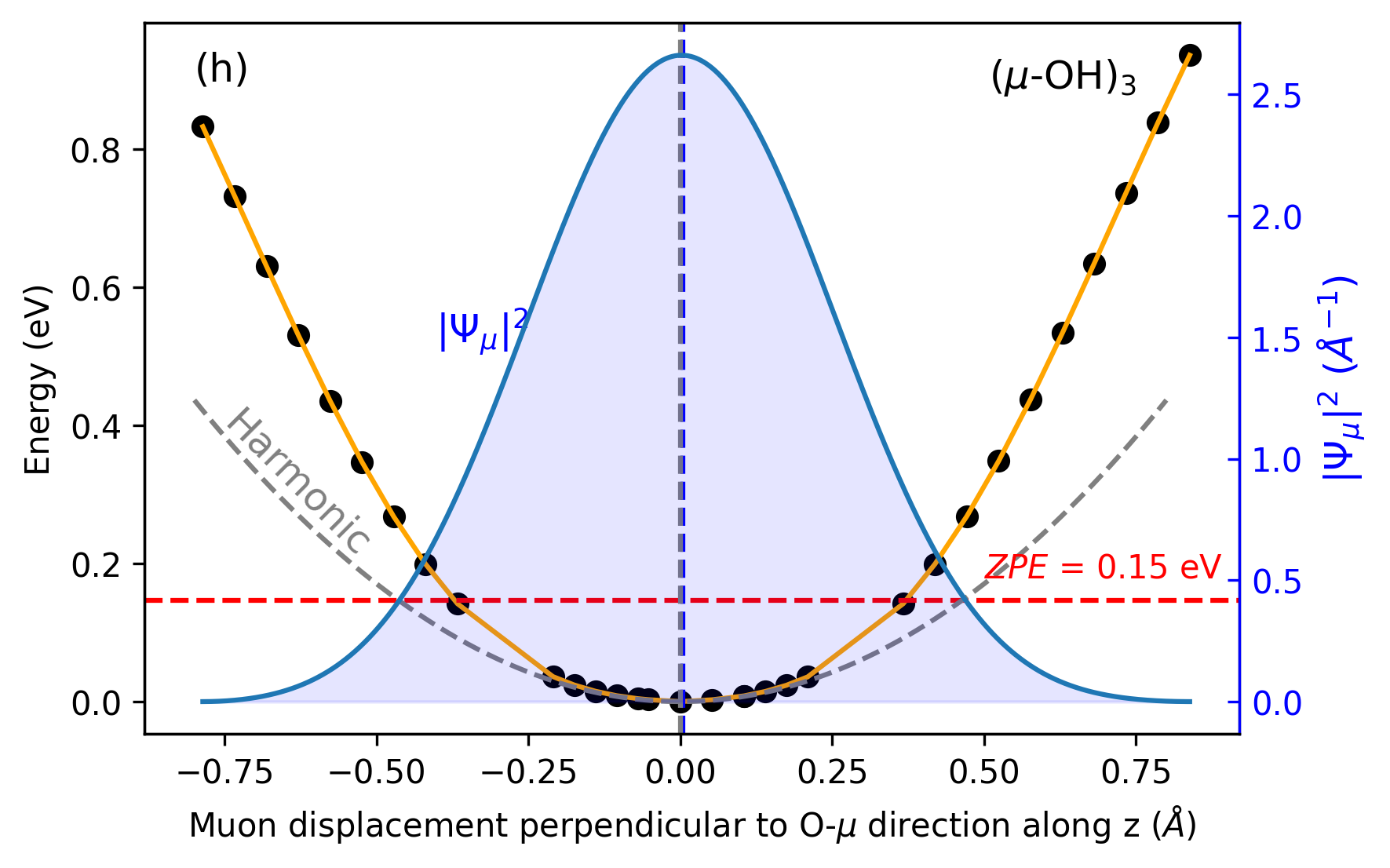}
    \label{ZPE_site4_Z}
  \end{subfigure}
  \caption{Effective potentials in the weakly-bound adiabatic limit (relative to the classical energy) for muon displacements perpendicular to (a--b) F--Br and (c--h) ($\mu$--O)$_i$ directions at F--$\mu$--Br and $\mu$--OH sites, respectively, from DFT (black points, interpolated by orange curves). These are compared to the effective potential under the harmonic approximation (dashed grey). Also shown here, the quantum muon probability distributions $|\psi_{\mu}|^{2}$ in the adiabatic potentials (blue curves) with the resulting average muon displacements ${\Delta}r_\mu$ (dashed vertical blue lines) away from their classical/harmonic positions (dashed vertical gray lines).}
  \label{harmonic}
\end{figure*}

 $\Gamma$-point phonon calculations were performed on a unit cell using  CASTEP \cite{ClarkSegallPickardHasnipProbertRefsonPayne+2005+567+570} with an energy cutoff of 2000 eV and a $4 \times 4 \times 3$ Monkhorst-Pack grid \cite{PhysRevB.13.5188} using the standard PBEsol functional \cite{PhysRevLett.100.136406} and a finite-displacement method. This allowed us to estimate the muon's quantum zero point energy (ZPE) in the harmonic approximation. Due to the low mass of the muon, the harmonic approximation underlying the phonon calculation is not \textit{a priori} justified. Therefore, the effective adiabatic potential in the weakly-bound limit $\cite{gomilvsek2023many}$ around the relaxed muon position was determined by sampling the muon energy at 35 different positions along the direction of the phonon normal modes of the muon. The adiabatic potential in the directions perpendicular to the main muon--ion direction ($\mu$--O or F--$\mu$--Br) are well described by the harmonic approximation and are summarized in Fig. \ref{harmonic}, while those along it are strongly anharmonic (see Fig. 3 in the main text).
 

\section{Details on the ${\mu}$SR experiments}
\label{Exp_conv_param}
The zero-field (ZF) and longitudinal-field (LF) $\mu$SR experiments were performed on the FLexible Advanced MuSR Environment (FLAME) instrument at PSI located at the $\pi$M3.3 beamline of the HIPA complex at the Swiss Muon Source (S$\mu$S) at PSI. The sample was pressed into a pellet with a $\approx$6 mm diameter, glued with GE-varnish to a $\approx$30 $\mu$m thick copper foil and mounted on a sample holder fork close to a cold finger to ensure good thermal conductivity. The magnetic field at the sample position was measured with a Hall probe, while active zero-field compensation using vector magnets reduced the field under zero-field conditions to < 1 $\mu$T. To be able to precisely measure the muon--nuclear spin oscillation, one has to collect much higher statistics than usual. For ZF measurement, 128 million muon/positron events were counted for the backward detector, and 127 million muons were in the forward detector. Between 25 million and 50 million events were gathered for the LF measurements. Data were fitted using the LMFIT Python package \cite{newville2016lmfit} using a microscopic approach outlined in \cite{lord2000muon,Wilkinson2020,celio1986new}. To properly treat the muon as a quantum mechanical particle, its dipolar Hamiltonian was averaged over the three-dimensional probability distribution of the muon $\psi_\mu|^2$, as in \cite{gomilvsek2023many,PhysRevMaterials.3.073804}. For each muon, the three-dimensional probability distribution consists of the three one-dimensional adiabatic potentials along the principal phonon modes of the muon (see previous section), and was numerically sampled using a  $30 \times 30 \times 30$ rectangular mesh. Due to computational limitations, the Hamiltonian was truncated to a dimension of 256, taking into account only interactions with the seven nuclei with spin $I > 0$ closest to the muon at a given muon site.
\end{document}